\documentclass{article}

\usepackage{arxiv}

\usepackage[utf8]{inputenc} % allow utf-8 input
\usepackage[T1]{fontenc}    % use 8-bit T1 fonts
\usepackage{hyperref}       % hyperlinks
\usepackage{url}            % simple URL typesetting
\usepackage{booktabs}       % professional-quality tables
\usepackage{amsfonts}       % blackboard math symbols
\usepackage{nicefrac}       % compact symbols for 1/2, etc.
\usepackage{microtype}      % microtypography
\usepackage{lipsum}		% Can be removed after putting your text content
\usepackage{graphicx}
\usepackage{natbib}
\usepackage{doi}

\newcommand{\vertexproperties}{P_v}
\newcommand{\edgeproperties}{P_e}

\usepackage{ifxetex}
\usepackage{ifluatex}
\newif\ifxetexorluatex % a new conditional starts as false
\ifnum 0\ifxetex 1\fi\ifluatex 1\fi>0
   \xetexorluatextrue
\fi

\ifxetexorluatex
  \usepackage{fontspec}
\else
  \usepackage[T1]{fontenc}
  \usepackage[utf8]{inputenc}
\fi

\usepackage{amsmath}
\usepackage{amssymb}
\usepackage{etoolbox}
\usepackage{xspace}
\newtoggle{textualoperators}

\usepackage{tikz}
\usepackage{forest}

\tikzset{every node/.style={draw}}

\newcommand{\tuple}[1]{\langle #1 \rangle}

\newcommand{\atom}[1]{\mathsf{#1}}

\newcommand{\colonseparator}{:}

\newcommand{\var}[1]{\mathtt{#1}}
\newcommand{\edgevariable}[2]{\var{#1}\ifstrempty{#2}{}{\colonseparator{\atom{#2}}}}

\newcommand{\kleenestar}{\ast}

\newcommand{\expandedgevariable}[4]{
	\left[
	\edgevariable{#1}{#2}
	\ifstrequal{#3}{1} % minHops = 1
	{
		\ifstrequal{#4}{1}
		{} % minHops = 1 and maxHops = 1 -> write nothing
		{\kleenestar_\atom{#3}^\atom{#4}} % minHops = 1 and maxHops != 1
	} % minHops != 1
	{\kleenestar_\atom{#3}^\atom{#4}}
	\right]}

\newcommand{\cartesianproduct}{\cartesianproduct}

\definecolor{red}{HTML}{e41a1c}
\definecolor{blue}{HTML}{377eb8}
\definecolor{green}{HTML}{4daf4a}
\definecolor{lilac}{HTML}{984ea3}

\definecolor{progressbargreen}{HTML}{008000}

\newcommand{\externalschemacolorname}{red}
\newcommand{\extravariablescolorname}{blue}
\newcommand{\internalschemacolorname}{green}

\colorlet{externalschemacolor}{\externalschemacolorname}
\colorlet{extravariablescolor}{\extravariablescolorname}
\colorlet{internalschemacolor}{\internalschemacolorname}

\usepackage{graphicx}
\usepackage{epsfig, wrapfig, multirow, enumitem, makecell, amssymb, cite}
\usepackage{tabu} % row font in table
\usepackage{hhline}
\usepackage{arydshln} % dash line table
\usepackage[ruled,vlined,linesnumbered]{algorithm2e}
\usepackage{graphicx}
\usepackage{tabulary}
\usepackage{booktabs}
\usepackage{multicol}
\usepackage{booktabs}
\usepackage[bb=boondox]{mathalfa}
\makeatletter
\DeclareRobustCommand\onedot{\futurelet\@let@token\@onedot}
\def\@onedot{\ifx\@let@token.\else.\null\fi\xspace}

\makeatother

\usepackage{listings}
\title{Rel2Graph: Automated Mapping From Relational Databases to a Unified Property Knowledge Graph}

\author{{\hspace{1mm}Ziyu Zhao} \\
	School of Physics, Mathematics and Computing\\
	The University of Western Australia\\
    Perth, Australia \\
	\texttt{ziyu.zhao@research.uwa.edu.au} \\
	\AND
	Wei Liu \\
	School of Physics, Mathematics and Computing\\
	The University of Western Australia\\
    Perth, Australia \\
	\texttt{wei.liu@uwa.edu.au} \\
	\And
	Tim French \\
    School of Physics, Mathematics and Computing\\
    The University of Western Australia \\
    Perth, Western Australia \\
	\texttt{tim.french@uwa.edu.au} \\
	\And
	Michael Stewart \\
    School of Physics, Mathematics and Computing\\
    The University of Western Australia \\
    Perth, Western Australia \\
	\texttt{michael.stewart@uwa.edu.au} \\
}

\renewcommand{\shorttitle}{\textit{arXiv} Template}

\hypersetup{
pdftitle={Rel2Graph},
pdfsubject={q-bio.NC, q-bio.QM},
pdfauthor={Ziyu Zhao, Wei Liu, Tim French, Michael Stewart},
pdfkeywords={Knowledge Graph Construction, Database migration, Query mapping.},
}

\begin{document}
\maketitle

\begin{abstract}
	Although a few approaches are proposed to convert relational databases to graphs, there is a genuine lack of systematic evaluation across a wider spectrum of databases. Recognising the important issue of query mapping, this paper proposes an approach Rel2Graph, an automatic knowledge graph construction (KGC) approach from an arbitrary number of relational databases. Our approach also supports the mapping of conjunctive SQL queries into pattern-based NoSQL queries. We evaluate our proposed approach on two widely used relational database-oriented datasets: Spider and KaggleDBQA benchmarks for semantic parsing. We employ the execution accuracy (EA) metric to quantify the proportion of results by executing the NoSQL queries on the property knowledge graph we construct that aligns with the results of SQL queries performed on relational databases. Consequently, the counterpart property knowledge graph of benchmarks with high accuracy and integrity can be ensured. The code and data will be publicly available. The code and data are available at github\footnote{https://github.com/nlp-tlp/Rel2Graph}.
\end{abstract}

% keywords can be removed
\keywords{Knowledge Graph Construction \and Database migration  \and Query mapping}
\section{Introduction}

\begin{figure*}[h!]
  \centering
    \scalebox{0.8}{\includegraphics[width=\textwidth]{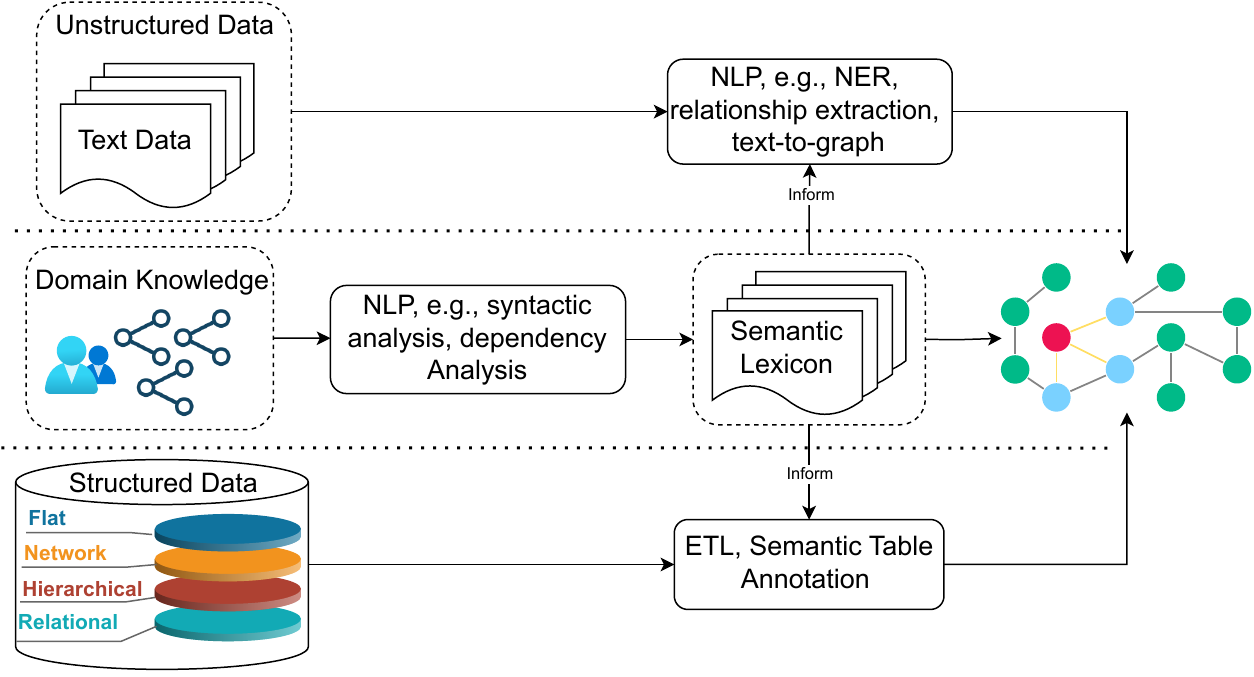}}
    \caption{The Taxonomy of Knowledge Graph Construction.}
    \label{fig:diagram}
\end{figure*}

A graph is a collection of interconnected triplets, where each triplet is composed of two nodes and an edge connected between them. Both nodes and edges can hold a set of attributes or properties in the form of key-value pairs~\citep{putrama2022automated}, such as a property graph which extends graphs by adding labels/types and properties for nodes and edges~\citep{marton2017formalising}. Knowledge Graph Construction (KGC)\footnote{Although knowledge graph is a broader concept that goes beyond a property graph, the techniques attached to KGC are also appliable to a property graph construction.} is an active research area aiming at representing knowledge from abundant data in a unified and standardized way. Any triple in a knowledge graph can be annotated with additional facts. It is a complex process to construct a knowledge graph from heterogeneous sources because it requires extracting and integrating information with respect to some uniform semantics. Heterogeneous sources range from unstructured data, such as plain text, to structured data, including tabular data such as CSVs and relational databases, and tree-structured formats, such as JSON and XML. We visualize the KGC process as a trio of paradigms shown in Figure~\ref{fig:diagram}. The uppermost paradigm embodies the knowledge extraction from unstructured data through a pipeline that encompasses natural language processing (NLP) tasks including named entity recognition (NER)~\citep{sang2003introduction}, entity linking~\citep{milne2008learning}, relation extraction~\citep{zelenko2003kernel} and event extraction~\citep{chen2015event}. The second one converges domain expertise to yield an informed knowledge graph such as Wikidata\footnote{https://query.wikidata.org/}, ConceptNet\footnote{https://conceptnet.io/}, and Bio2RDF\footnote{https://bio2rdf.org/sparql}. Notably, the semantic lexicon derived from domain expertise assumes a pivotal role in annotating the extracted triplets, thereby enhancing their contextual significance. The third paradigm is characterized by the transformation of structured data such as tabular data into a knowledge graph. This transformation can be achieved through the Extract, Transform, and Load (ETL) process. An alternative avenue involves semantic table annotation~\citep{abdelmageed2020jentab}. 

In this paper, we align with the ETL paradigm, harnessing its streamlined mechanism to automatically construct a property knowledge graph in Neo4j\footnote{neo4j.com/} from arbitrary relational databases (Section 4.1). A relational database is a structured collection of tabular data. Tabular data refers to data that is organized into a two-dimensional structure with rows and columns, i.e. a table. Each table represents a specific entity or concept, and the relationships between these entities are established through primary keys and foreign key constraints. The resulting knowledge graph from relational databases can be efficiently utilized in a variety of downstream NLP tasks, including semantic parsing~\citep{ sorokin2021knowledge}, natural language interface~\citep{zhao2022natural}, and question answering~\citep{chakraborty2019introduction}. Among these, semantic parsing plays an essential role and it aims to translate a natural language utterance into a formal query language, such as Text-to-SQL and Text-to-Cypher. Consequently, we also address the challenge of query mapping, specifically focusing on the translation of SQL into Cypher queries (Section 4.2). This undertaking is critical for aligning the application layer, which deploys semantic parsing, with the new persistent layer, ensuring optimal efficiency. In this context, we repurpose well-established semantic parsing benchmarks, originally designed for relational databases, such as Spider~\citep{yu2018spider} and KaggleDBQA~\citep{lee-2021-kaggle-dbqa}, to propel advancements in the retrieval of large graph-based knowledge repositories.

\begin{figure*}[h]
  \centering
    \scalebox{0.9}{\includegraphics[width=\textwidth]{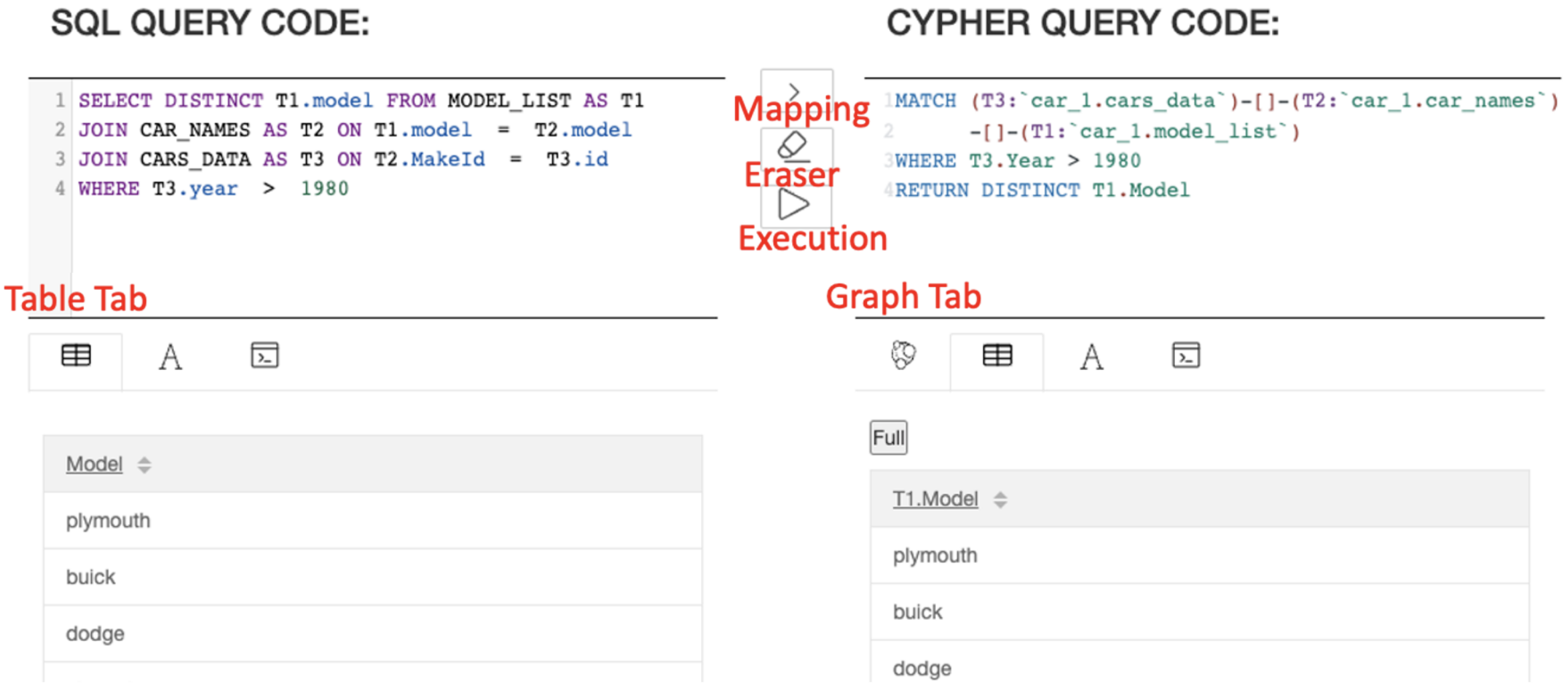}}
     \caption[]{The web-based interface visualises the conjunctive query mapping process of our proposed {\tt Rel2Graph}. It enables users to compose an SQL query and obtain the semantically equivalent Cypher query. }
    \label{fig:interface}
\end{figure*}

With respect to querying relational databases, the challenge becomes particularly pronounced when dealing with intricate inter-table relationships. The inherent complexity arises from the rich relational structure of entities within complex relational databases. This complexity is most evident in the endeavor to craft a target SQL query that encompasses multiple join operations. In contrast, a property graph naturally accommodates interconnected tabular data due to its explicit representation of relationships through edges between nodes. Figure~\ref{fig:interface} displays our query mapping process\footnote{The Web UI is adapted from https://github.com/UNSW-database/SQL2Cypher.}. The initial query instance represents a sophisticated SQL query featuring two {\tt JOIN ON} statements. However, once our query mapping rules are applied, the resulting Cypher query that maintains semantic equivalence no longer necessitates resource-intensive join operations. The divergences between SQL and Cypher are discussed in Section 3.

In this endeavor, we introduce an approach {\tt Rel2Graph} to construct a property knowledge graph automatically from relational databases schema dump. The {\tt Rel2Graph} approach also supports the mapping of conjunctive SQL queries into pattern-based Cypher queries. We evaluate our proposed approach using the widely-adopted Spider~\citep{yu2018spider} and KaggleDBQA~\citep{lee-2021-kaggle-dbqa} benchmarks. Spider is a large-scale dataset. It contains 5,693 unique complex SQL queries on 200 databases with multiple tables, covering 138 different domains. KaggleDBQA is a relatively new cross-domain evaluation dataset of real Web databases. It includes real-world databases from Kaggle\footnote{https://www.kaggle.com}, a platform for data science competitions and dataset distribution. The relational databases in KaggleDBQA feature abbreviated and obscure column names, domain-specific categorical values, and minimal preprocessing. We employ execution accuracy (EA)~\citep{LinRX2020:BRIDGE} metric to gauge data consistency at the semantic level. EA quantifies the alignment between the results of SQL queries executed on relational databases and those resulting from graph-based queries performed on the generated property knowledge graph. Consequently, Rel2Graph empowers the utilization of inherent structured data within relational databases for constructing a path-pattern-focused property knowledge graph, highly relevant to semantic parsing task.

The rest of the paper is organised as follows. Section 2 reviews the related work on the KGC process. Section 4 introduces the {\tt Rel2Graph} approach in detail. Section 5 discusses the conversion result and error analysis. Section 6 is the conclusion. 

\section{Related Work}
In this paper, we review the following two paradigms that are used in knowledge graph construction (KGC) tasks, i.e. KGC from unstructured text and KGC from structured data.

\subsection{KGC from Text}
Knowledge graph construction from unstructured text aims to extract structural information organized in a triplet format. Generally, there are discriminative and generative methodologies for KGC from text data~\citep{ye2022generative}.

\paragraph{Discriminative Methodology}
The goal is to predict the labels given the input text for each element in a triplet which is then populated into a knowledge graph. For instance, given an input text in maintenance domain \textit{oil leak feed pump}, a NER model identifies the named entity \textit{feed pump} and tags it using {\tt /Item} label, the same as predicting {\tt /Observation} label for \textit{leak}, in order to generate a triplet {\tt </Item, has\_observation, /Observation>}~\citep{stewart2022mwo2kg}. The prediction label set is typically constructed by domain experts. Apart from NER~\citep{chiu2016named}, the discriminative models consist of other natural language processing (NLP) tasks, e.g., Entity Linking~\citep{paolini2021structured}, Relation Extraction (RE)~\citep{wei2019novel} and Event Extraction~\citep{chen2015event}. 

A high-quality annotated text corpus should be ready for the discriminative models. There are a lot of collaborative annotation tools, such as BRAT~\citep{stenetorp-etal-2012-brat}, Doccano~\citep{doccano}, Redcoat~\citep{stewart-etal-2019-redcoat} and QuickGraph~\citep{bikaun-etal-2022-quickgraph}, that have been proposed to produce high-quality training datasets for machine learning practitioners. A team of 12 workers has employed Redcoat~\citep{stewart-etal-2019-redcoat} to carry out collaborative hierarchical entity annotation on maintenance work orders (MWOs), and they use the annotated corpus to train a maintenance NER model. Echidna~\citep{stewart2022mwo2kg} then uses the pre-trained NER model\footnote{https://huggingface.co/nlp-tlp/mwo-ner} to detect named entities tagged by the pre-define entity label set as the graph nodes towards constructing a maintenance knowledge graph. During the MWO2KG process, a maintenance asset hierarchy taxonomy has been applied to augment the coarse entity label set to a fine-grained label set, for example, the named entity {\tt pump} (as a graph node) tagged by the coarse label {\tt /Item} is further assigned the fine-grained label set [{\tt /rotating equipment/Item, /Item}] in the sense of enriching maintenance knowledge graph expressiveness. Overall, the discriminative approach to KGC is achieved through the utilization of high-quality training data created by domain experts. However, acquiring high-quality annotated corpora for complex multi-task information extraction is an arduous and costly process for human annotators, and it relies heavily on specialized domain knowledge either stored in dictionaries, gazetteers, or ontology format. Fortunately, there are some available annotation tools, such as QuickGraph~\citep{bikaun-etal-2022-quickgraph}, designed to allow domain experts to annotate data for the full pipeline of KGC from text, i.e. NER and RE.

\paragraph{Generative Methodology}
For the automatic generative methodology to KGC, the goal is to autoregressively generate linearized triplets given the input sentence by fine-tuning sequence-to-sequence (seq2seq) models. For example, BT5~\citep{agarwal2020machine} proposed to use T5~\citep{raffel2020exploring} to generate KG in a linearized form. There are a number of proposed research works aiming to generate the graph structure ground up. DualTKB~\citep{dognin2020dualtkb} explored GRU~\citep{chung2014empirical}, Transformer~\citep{vaswani2017attention} and BERT~\citep{devlin2018bert} in an encoder-decoder architecture to enable the text-to-graph translation in both directions using an unsupervised cycle loss, i.e. the back-translations of a sample from a generated triplet back to itself. 

The potential issue with seq2seq or end-to-end modeling is that the graph linearisation is not unique and inefficient due to the repetition of graph components multiple times~\citep{melnyk2022knowledge}. GraphRNN~\citep{you2018graphrnn} decomposes the graph generation process into a sequence of node and edge formations using two RNNs, conditioned on the graph structure. CycleGT~\citep{guo2020cyclegt} undertakes text-to-graph in an unsupervised way using an entity extractor followed by a relationship classifier on top of T5. Grapher~\citep{melnyk2022knowledge} is a two-stage KG generation, namely the graph nodes (entities) generation process using T5 and the graph edges generation using the available entity information. 

While the idea of formulating KGC from the text in the sequence-to-sequence fashion is free from the constraints of dedicated architectures, expensive labour annotation exercises, and specialized knowledge sources, the sequential and greedy nature of its generation can cause sub-optimal graph structure. It lacks efficient architectures specialized for graph-structured generation output and limited
parallel training data. Hence, it is more controllable and efficient to construct large-scale KG through the elaborate discriminative methods in a pipeline strategy to solve a specific task. Nevertheless, the pre-trained language models, such as T5~\citep{raffel2020exploring} and BART~\citep{lewis2019bart} that store the vast amounts of linguistic knowledge can be exploited for the downstream NLP tasks of domain-specific KG construction from text data in a unified multi-task learning setting.

\subsection{KGC from Structured Data}
Structured data sources play a fundamental role in the data ecosystem because much of the valuable and reusable information within enterprises and the Web is available as structured data. 

\paragraph{ETL Process}

Extract, transform, and load (ETL) is the process of combining structured data from multiple sources into a unified data warehouse, e.g., a graph database. It originated with the emergence of relational databases to convert transactional databases to relational databases. 

Numerous commercial enterprises have developed ETL tools, exemplified by the Neo4j ETL tool\footnote{https://neo4j.com/labs/etl-tool/}. This tool facilitates the importation of data from relational databases into property graph databases hosted within Neo4j. While the Neo4j ETL tool effectively handles basic ETL processes, it may encounter limitations when confronted with exceptionally intricate transformations and tasks involving data cleansing or normalization. In contrast, our proposed {\tt Rel2Graph} approach demonstrates enhanced efficiency in managing these complex tasks. Moreover, the performance of the Neo4j ETL tool may suffer when confronted with substantial datasets or intricate transformation requirements. Additionally, our proposed approach supports customized query mapping, varying from simple queries to complex queries, on top of data migration from large-scale relational databases to property graph databases. This flexibility extends to providing options for data enrichment, encompassing capabilities such as semantic parsing and question answering (KGQA).

For the Semantic Web community, a common strategy involves adopting reference ontologies as global schemas through semantic integration processes that map structured information onto Knowledge Graphs (KGs). Table2Graph~\citep{lee2015table2graph} has been employed to construct graphs based on the Map-Reduce framework over Hadoop, drawing data from three distinct healthcare databases. MDM2G~\citep{lachicheperformance} has proposed a set of formal rules enabling the conventional star and snowflake schemas to fit in the ETL process from a relational database to a graph database. 

In the context of this paper, the most closely related works are those proposed by Putrama et al.~\citep{putrama2022automated} and Feng et al.~\citep{feng2022approach}. The former evaluated the automated graph construction from four relational databases to Neo4j, primarily comparing the total number of records in the source databases to the nodes and edges created in the resulting graph database. On the other hand, the latter evaluated the data consistency using relatively straightforward simple queries but with less efficiency. By contrast, our study undertakes a thorough assessment of automatic database transformation across two distinct benchmarks: Spider and KaggleDBQA. This evaluation encompasses significantly more intricate applications, covering large-scale relational databases. Additionally, it extends to a comprehensive examination of data consistency at the semantic query level.

\section{Preliminary - SQL v.s. Cypher}
The proliferation of graph databases, driven by the increasing complexity and interconnectedness of data across various domains, has created a pressing need for intuitive and accessible methods to interact with these databases in many real-world scenarios. Property graph databases, for example, Neo4j, Azure DB, OrientDB, and ArangoDB, extend graph databases by adding not only semantic labels to graph nodes and types to graph edges but also properties on both nodes and edges. These additional features enrich a domain data model, providing more context and flexibility to represent real-world applications, such as recommendation systems~\citep{wang2019kgat} and fraud detection~\citep{pourhabibi2020fraud}. Property graph databases, in conjunction with declarative query languages, offer a robust solution for storing, retrieving, and analyzing data encompassing intricate graph patterns. This approach proves particularly advantageous when compared to conventional relational databases, as the latter would present challenges in terms of both efficiency and manageability when dealing with such complex graph structures~\citep{szarnyas2018train}. Traditional SQL-based database systems (aka relational databases) have long been the dominant choice for data storage and querying, however, they are not always well-suited for handling highly connected data.

Neo4j, topping the DB engine ranking of GraphDBMS\footnote{https://db-engines.com/en/ranking}, is a popular industry scale non-relational (NoSQL) property graph database. It offers high-level Cypher query language to specify graph patterns supported by the query engine that uses sophisticated optimization techniques~\citep{marton2017formalising}. SQL uses the syntax with {\tt SELECT}, {\tt FROM}, {\tt JOIN}, {\tt GROUP BY}, and {\tt HAVING} clauses, while Cypher employs a more pattern-based syntax with {\tt MATCH}, {\tt WHERE}, {\tt WITH}, and {\tt RETURN} clauses. Patterns and pattern-matching are the core building blocks of Cypher queries. A pattern describes the data using nodes, relationships between any two nodes, and their properties. It is very similar to how one typically draws the shape of property graph data on a whiteboard, and intuitively it can better model complex relationships and connected data. Patterns appear in multiple places in Cypher, such as {\tt MATCH} and {\tt WHERE} as shown below. Note that, we only describe the basic and frequent usage of Cypher patterns\footnote{For more details, please refer to Cypher query language reference.}. 

\begin{enumerate}
    \item  
    Match a graph node with a specific label using parentheses, e.g., {\tt MATCH (T1:department)}, where {\tt department} is the node label and {\tt T1} is the alias. 
    \item 
    Match a graph edge with a type between two nodes such as {\tt MATCH (T1:department)-[T2:management]-()}.
    \item  
    Match a pattern alongside specifying properties using curly brackets surrounding a number of key-expression pairs, separated by commas, e.g., {\tt MATCH (T1:department)-[T2:management \{temporary\_acting:`Yes'\}]-()}, where {\tt temporary\_acting} is the property of the edge with value `Yes'. The pattern is to identify the departments managed by heads who are temporarily acting.
    \item 
    Using path patterns in {\tt WHERE} by matching on a relationship type using square brackets and either directed or undirected arrows. The symbol {\tt -[]-} means related to, without regard to the type or direction of the relationship, e.g., {\tt WHERE ()-[T2:management]-()}.

\end{enumerate}

\section{Automated Approach Overview}

\paragraph{\bf Problem Formulation} Given a source dataset denoted as $D_R=(\mathcal{S}, Q_{sql})$, we aim to construct $D_G=(\mathcal{G}, Q_{cyp})$. $\mathcal{S}$ represents a collection of relational databases, and $\mathcal{G}$ denotes a corresponding property knowledge graph. $Q_{sql}$ signifies a set of SQL queries and $Q_{cyp}$ denotes the respective set of Cypher queries. 

The objective involves a twofold process, i.e. database mapping ($\mathcal{S} \xrightarrow{\text{map}} \mathcal{G}$) and query mapping ($Q_{sql} \xrightarrow{\text{map}} Q_{cyp}$). Firstly, we incrementally construct the property knowledge graph $\mathcal{G}$ from the relational databases in $\mathcal{S}$ using Algorithm 1 (Section 4.1). Secondly, we translate SQL queries to Cypher queries (Section 4.2). To evaluate the quality of our database and query mappings, we assess data consistency. Specifically, we measure how accurately the Cypher queries $Q_{cyp}$ executed against $\mathcal{G}$ align with the source SQL queries $Q_{sql}$ over $\mathcal{S}$. The execution accuracy (EA) metric is employed to quantify this alignment (Section 5).

\begin{figure*}[h!]
  \centering
    \scalebox{0.9}{\includegraphics[width=\textwidth]{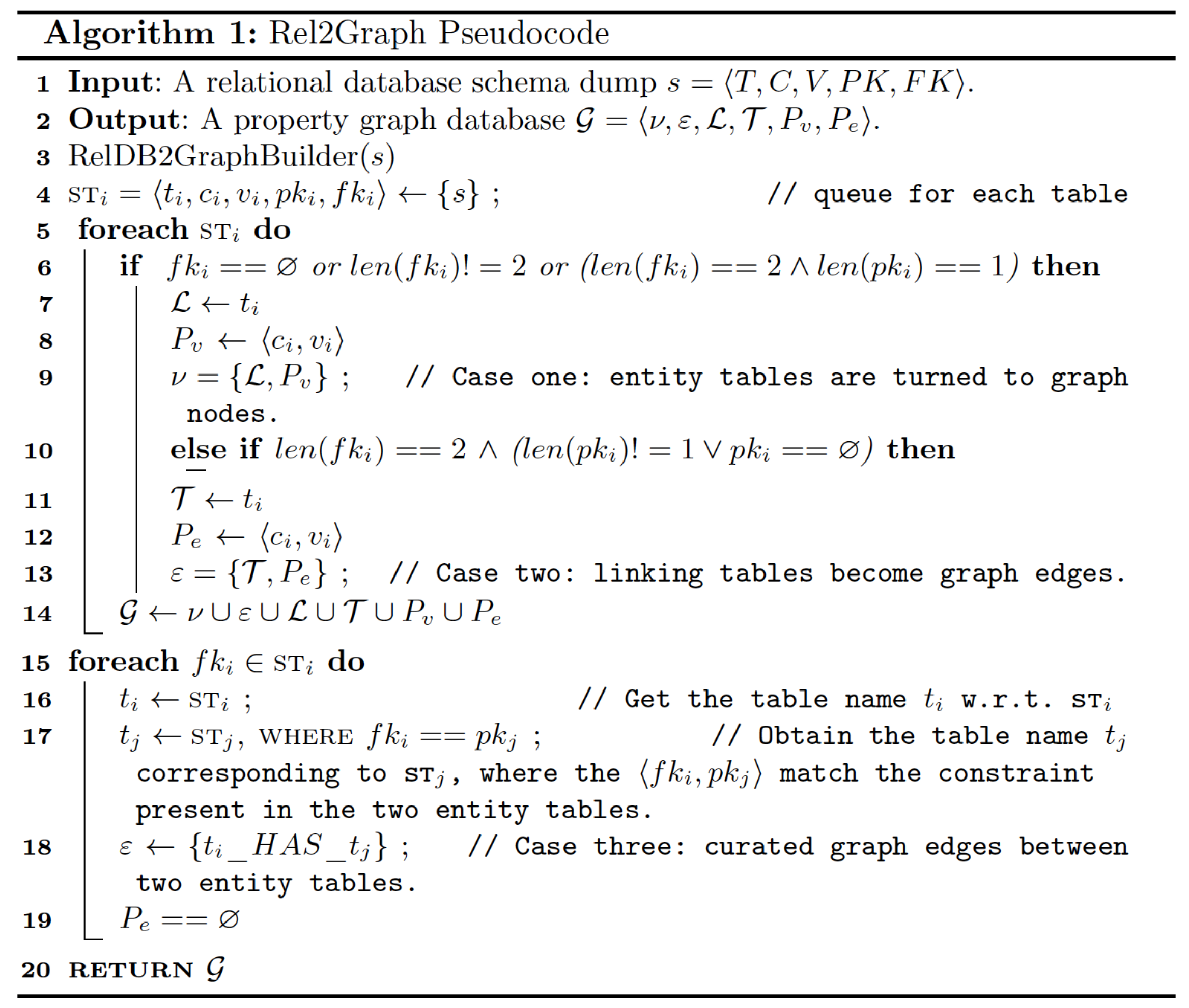}}
    \label{fig:algorithm}
\end{figure*}

Continuing the notation introduced in Algorithm 1, we symbolize each relational database within $\mathcal{S}$ as $s = \langle T, C, V, PK, FK \rangle$, where the elements of the tuple correspond to collections of table names, column names, row values, primary keys, and foreign keys. We incrementally convert $s$ into the targeted property knowledge graph $\mathcal{G} = \tuple{\nu, \varepsilon, \mathcal{L}, \mathcal{T}, \vertexproperties, \edgeproperties}$, where the elements of the tuple represent graph nodes, graph edges, node labels, edge types, and the properties of nodes and edges, respectively. We illustrate the mapping process following.

\subsection{Database Mapping}

\paragraph{\bf Metadata Analyzer}
To ensure comprehensive consideration of various table types and their inherent relationships within the database mapping procedure automatically, we recognize the significance of primary and foreign key constraints. This leads us to define the automated examination of these constraints as a metadata analyzer. Each pair of primary and foreign key constraint is represented as $\langle pk, fk\rangle$, where $pk \in PK$ and $fk \in FK$, for the purpose of distinguishing between entity tables and linking tables (or associative entity tables).  

To recognize an entity table within a relational database, several key criteria, including primary keys and foreign keys, play a crucial role. Entity tables often exhibit specific characteristics that set them apart. For instance, they commonly possess a primary key designed to uniquely identify each entity or record. When a table features a primary key composed of a single column, it becomes a strong candidate for representing an entity. Additionally, entity tables tend to be less reliant on relationships with other tables. This is evident when a table lacks foreign keys, which can indicate its status as an entity table. Besides, entity tables tend to have one-to-many or one-to-one relationships with other tables. Tables that participate in many-to-many relationships are less likely to be entity tables. Apart from primary keys, foreign keys, and relationship cardinality, which can be queried using a relational database engine, our analysis includes the utilization of available schema documentation or data dictionaries (e.g., \textit{KaggleDBQA\_tables.json}) when migrating KaggleDBQA databases into corresponding property graphs. This documentation explicitly outlines the primary keys and foreign keys associated with related tables, offering extra information in cases where querying primary key and foreign key constraints yields no results.

It is important to acknowledge that while primary keys, foreign keys, and relationship cardinality are important factors, other criteria, such as the semantic meaning of a table, naming conventions, and query analysis, may also come into play when identifying entity tables in different contexts or methodologies. However, in order to provide a structured and automated analysis for identifying entity tables within the context of our metadata analyzer, we ensure that a table qualifies as an entity table if it satisfies any of the following three conditions.

\begin{enumerate}
    \item A table is devoid of any foreign key, i.e. $fk == \varnothing$.
    \item A table either encompasses a sole foreign key or involves more than two foreign keys, i.e. the count of foreign keys ($len(fk)$) is not equal to 2.
    \item A table possesses two foreign keys and a primary key consisting of just one column, indicated by the condition $len(fk)==2 \land len(pk)==1$.
\end{enumerate} 

Subsequently, these entity tables are translated into graph nodes, where each node is labelled with the corresponding entity table name. Each row within an entity table corresponds to a node, while the columns in entity tables are transformed into properties of the nodes. This particular scenario is elaborated upon from line 6 to line 9 in Algorithm 1. These entity tables serve as the foundation for constructing a knowledge graph or property graph.

Secondly, we argue a table becomes a linking table if the number of foreign keys ($len(fk)$) is equal to 2, and meanwhile, one of the two additional conditions holds, i.e. the number of primary keys ($len(pk)$) is not equal to 1, or $pk== \varnothing$. Linking tables become graph edges. Each table name is represented by a type on the edges. Each row in a linking table is an edge. Columns of linking tables become edge properties. For further details, refer to the pseudocode outlined from line 10 to line 13.

\begin{figure*}[h]
% \vspace{-1em}
  \centering
    \scalebox{1}{\includegraphics[width=\textwidth]{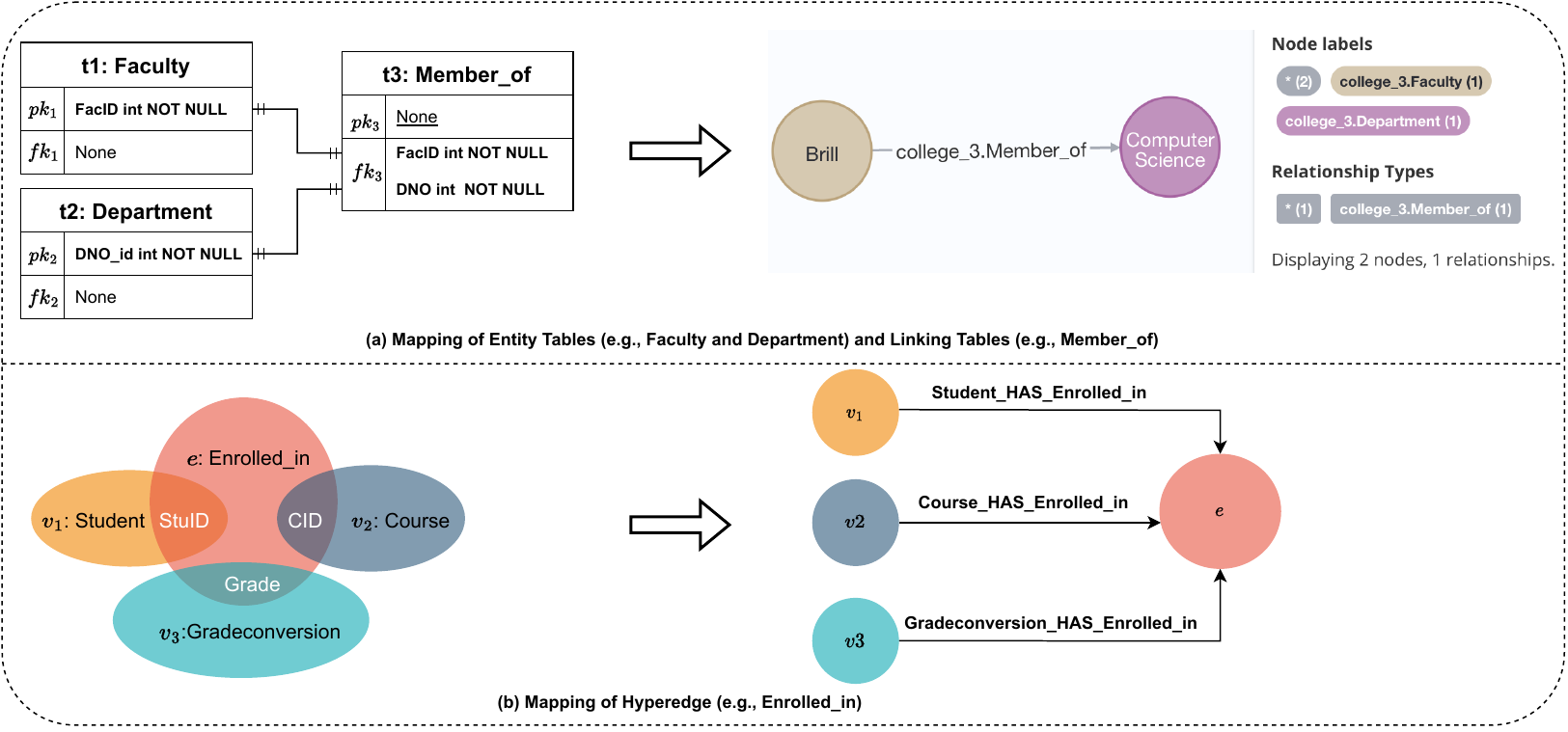}}
     \caption[]{An illustration showcasing the handling of three mapping scenarios. These include: a) mapping entity tables such as {\tt Faculty} and {\tt Department} into graph nodes labelled with their respective table names, and mapping the linking table {\tt Member\_of} into graph edges, typed by the table name, and b) the transformation of a hyperedge represented as {\tt e}.}
    \label{fig:rel2graph-example}
% \vspace{-1em}
\end{figure*}

In addition to the aforementioned scenarios, when neither of the two arguments is met, we establish graph edges typed in the format of {*\_HAS\_*} between two entity tables. As illustrated in Figure~\ref{fig:rel2graph-example}, we generate a graph edge typed as {\bf Student\_HAS\_Enrolled\_in} connecting graph nodes labelled as {\bf Student} and {\bf Enrolled\_in}, respectively. 

Based on our metadata analyzer, as demonstrated in Figure~\ref{fig:rel2graph-example}, the entity-relationship (ER) diagram located in the top left illustrates the relational database \textsc{colledge\_3.sqlite}. It includes an entity table labeled as $t_1 = {\bf Faculty}$ with its primary key $pk_1 =\tuple{  \textsc{FacID}}$, an entity table $t_2 = {\bf Department}$ with $pk_2 = \tuple{\textsc{DNO}}$, and an linking table $t_3 = {\bf Member\_of}$ alongside two foreign keys $fk_3 =\tuple{\textsc{FacID}, \textsc{DNO}}$. The connection between $t_1$ and $t_3$ is established via the primary and foreign key constraint $\tuple{pk_1, fk_3}$ where the $ pk_1 \subset fk_3$ holds. Therefore, the linking table $t_3$ is mapped to graph edges, connecting the graph nodes converted from the entity tables $t_1$ and $t_2$. An instance of such mapping is depicted in the top right of Figure~\ref{fig:rel2graph-example}.

What's more, illustrated in the bottom left, we offer an example of a hyperedge $e$ typed as {\bf Enrolled\_in}. The concept of a hyperedge arises from graph theory to represent relationships that involve more than two graph nodes. A hyperedge is useful when dealing with relationships that do not fit neatly into a one-to-one connection but involve multiple entities interacting together. Nevertheless, due to the limitation of the Neo4j data model, we can not visualise such hyperedges. Consequently, we adopt an alternative approach by turning the hyperedge $e$ into graph nodes and apply the mapping rule outlined in the last scenario to curate three distinct graph edges connecting the transformed hyperedge $e$ to the nodes $v_1$, $v_2$, and $v_3$ correspondingly. This adaptation allows us to capture the essence of hyperedge relationships within the constraints of the visualization framework.

\paragraph{\bf Data Repairment}

During the database mapping procedure, the presence of data issues, for example, inconsistencies in data types of columns working as primary keys and foreign keys, can lead to errors and mismatching issues during the query mapping procedure in Section 4.2. We summarize and address four main data issues identified by our metadata analyzer as follows.

1) {\bf No Primary Key}. Within the Spider database, 95 out of the 876 tables (approximately 10.8\%) do not feature a primary key. The approach to addressing the absence of a primary key in a relational database can be highly context-dependent, particularly during the database mapping process. In our specific case, we have identified that the absence of a primary key would result in the inability to recognize both primary and foreign key constraints. To mitigate this issue, we opt to designate a column or combination of columns that are already functioning as foreign keys in other related tables as the primary keys for these tables. In situations where no suitable candidate exists, we choose to leave the table as-is, maintaining its status without a primary key.

2) {\bf No/Incorrect Foreign Key}. In the KaggleDBQA dataset, explicit foreign keys are not defined within the \textsc{StudentMathScore.SQLITE} database, neither in the schema dump nor mentioned in the \textit{KaggleDBQA\_tables.json} file. However, we observed that out of the total twenty-eight provided SQL queries, twenty-one of them involve the utilization of {\tt JOIN ON} statements. In these instances, specific column names are employed in a manner reminiscent of foreign keys within the SQL {\tt JOIN ON} statements. For instance, we can deduce the presence of a foreign key relationship involving {\tt state\_code} from the following statement:

\begin{lstlisting}
FINREV_FED_17 as T1 JOIN FINREV_FED_KEY_17 as T2 
ON T1.state_code = T2.state_code
\end{lstlisting}

Similarly, we can infer the existence of a foreign key association involving {\tt state} from this statement:
\begin{lstlisting}
FINREV_FED_KEY_17 as T2 JOIN NDECoreExcel_Math_Grade8 as T3
ON T2.state = T3.state
\end{lstlisting}

To address this situation and establish meaningful graph edges within the SQLite databases that lack explicitly defined foreign key constraints, we propose a solution. This solution leverages the column names utilized in SQL {\tt JOIN ON} statements as implicit associations, allowing us to create coherent graph relationships.

In the Spider dataset, the error in the provided SQL {\tt CREATE} statements of \textsc{musical.SQLITE} is in the foreign key constraint definition of the {\tt actor} table. Specifically, the FOREIGN KEY clause references the {\tt actor} table itself rather than referencing the {\tt musical} table, which is logically incorrect and can lead to database integrity issues. To correct the error, the FOREIGN KEY clause in the {\tt actor} table should be modified to correctly reference the {\tt musical} table and its primary key.

\begin{lstlisting}
CREATE TABLE musical (
Musical_ID int,
Name text,
Year int,
Award text,
Category text,
Nominee text,
Result text,
PRIMARY KEY (Musical_ID)
);

CREATE TABLE actor (
Actor_ID int,
Name text,
Musical_ID int,
Character text,
Duration text,
age int,
PRIMARY KEY (Actor_ID),
FOREIGN KEY (Musical_ID) REFERENCES actor(Actor_ID)
);
\end{lstlisting}

As a result, we verify and maintain the consistency of foreign keys within the database mapping process. Above all, we take the precaution of excluding foreign keys from the set of properties associated with an edge. This helps prevent potential data consistency issues when altering the related primary keys of nodes.

3) {\bf Database content related Issue}. In the Spider dataset, out of the 876 tables, 11 tables (1.3\%) contain duplicate rows. During the graph construction process, all duplicate rows are automatically eliminated, although this may lead to discrepancies when calculating the execution accuracy score. Besides, there are seven relational databases, each of which contains only table columns but no table content. In these cases, we assign an empty value to each table column to avoid collapsed database migration. 

4) {\bf Graph Node Label Differentiation Issue}. One of the drawbacks of Neo4j is its inability to differentiate between graph node labels. For instance, consider two tables both named {\tt singer} in \textsc{singer.sqlite} and \textsc{concert\_singer.sqlite}. If we were to store both databases in the same graph database without reformatting the table names, it could lead to a decrease in the execution accuracy score.

As depicted in Table~\ref{tab:error_analysis}, even when the generated Cypher query is correct, the EA may not exactly match the gold SQL execution answer. A closer examination reveals that the discrepancy, such as the case of {\tt Justin Brown}, belongs to \textsc{concert\_singer.sqlite} rather than \textsc{singer.sqlite}. This error can be rectified by renaming each table using the format {\tt domain\_name.table\_name}.

\begin{table*}[h!]
    \centering
    \resizebox{\textwidth}{!}
    {\begin{tabular}{l|l}
    \toprule
   Example 1 & \bf A nested query example. \\
    \toprule
    NLQ: & {\tt What is the name of every singer that does not have any song?}\\
    SQL query: & {\tt SELECT Name FROM singer WHERE Singer\_ID NOT IN (SELECT Singer\_ID FROM song) } \\
    SQL query ER: & \bf [[`Alice Walton'], [`Abigail Johnson']]\\
    Cypher query: & {\tt MATCH (si:singer) WHERE NOT (si:singer)-[]-(:song) RETURN si.Name} \\ 
    Cypher query ER: & \bf [[`Justin Brown'], [`Alice Walton'], [`Abigail Johnson']]\\
    \midrule
   Example 2 & \bf An example that duplicate elements are dropped automatically. \\
    \midrule
    NLQ: & {\tt Which skill is used in fixing the most number of faults? List the skill id and description.?}\\
    SQL query: & {\tt SELECT T1.skill\_id, T1.skill\_description FROM Skills AS T1 JOIN Skills\_Required\_To\_Fix} \\
    & {\tt AS T2 ON T1.skill\_id  =  T2.skill\_id GROUP BY T1.skill\_id ORDER BY count(*) DESC LIMIT 1} \\
    SQL query ER: &  \bf [[3, `TV, Video']]\\
    Cypher query: & {\tt MATCH (T1:`assets\_maintenance.Skills`)-[T2:`assets\_maintenance.Skills\_Required\_To\_Fix`]-()} \\
    & {\tt WITH T1.skill\_description AS skill\_description, count(T1.skill\_id) AS count, T1.skill\_id AS skill\_id }\\
    &{\tt RETURN skill\_id,skill\_description ORDER BY count DESC, LIMIT 1}\\ 
    Cypher query ER: & \bf [[2, `Mechanical']]\\
    \bottomrule
    \end{tabular}}
    \caption[]{The examples for error analysis. {\tt ER} is short for execution results.}
    \label{tab:error_analysis}
\end{table*}

5) {\bf Others}. While the above issues represent the main challenges we address during the graph construction process, it is important to acknowledge that other conditions for data repair may exist in different contexts. These could include issues related to data formatting issues such as the semantic meaning of a table and column naming conventions. Our approach is designed to provide a structured and automated analysis for identifying and repairing data issues within the context of our metadata analyzer. These conditions ensure that tables are categorized accurately based on their relational structure.

\subsection{Query Mapping}

\begin{figure*}[h!]
  \centering
    \scalebox{1}{\includegraphics[width=\textwidth]{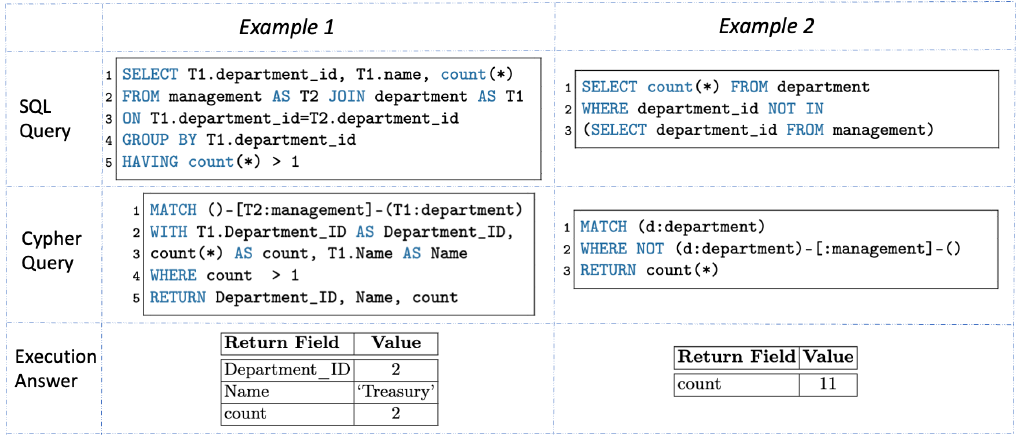}}
    \caption{The goal SQL queries (top) of example 1 and example 2, their respective Cypher queries (middle) and the Cypher queries execution answers (bottom).}
    \label{fig:examples}
\end{figure*}

Graph data models are known for their efficient pattern-matching mechanisms in dealing with highly connected data when relational databases resort to multiple expensive {\tt JOIN ON} operations via foreign keys. Take the two examples in Figure~\ref{fig:examples} for instance. In Example 1, \textit{ Which department has more than 1 head? List the id, name, and number of heads}, we can directly map the {\tt JOIN ON} statement into a Cypher pattern in {\tt MATCH} clause statement where the foreign key \textit{ department\_id} is omitted. For the question of Example 2, \textit{ How many departments are led by heads who are not mentioned}, we can convert the nested sub-query into a graph pattern in the {\tt WHERE} statement. The equivalent Cypher clauses of both examples do not need to explicitly mention the foreign key \textit{department\_id}. Foreign keys become redundant in graph databases. Due to the fact that Cypher queries are case-sensitive, we normalised all the schema items appearing in the source case-insensitive SQL queries by using the graph schema which is aligned with relational databases. 

\begin{figure*}[h!]
  \centering
    \scalebox{1}{\includegraphics[width=\textwidth]{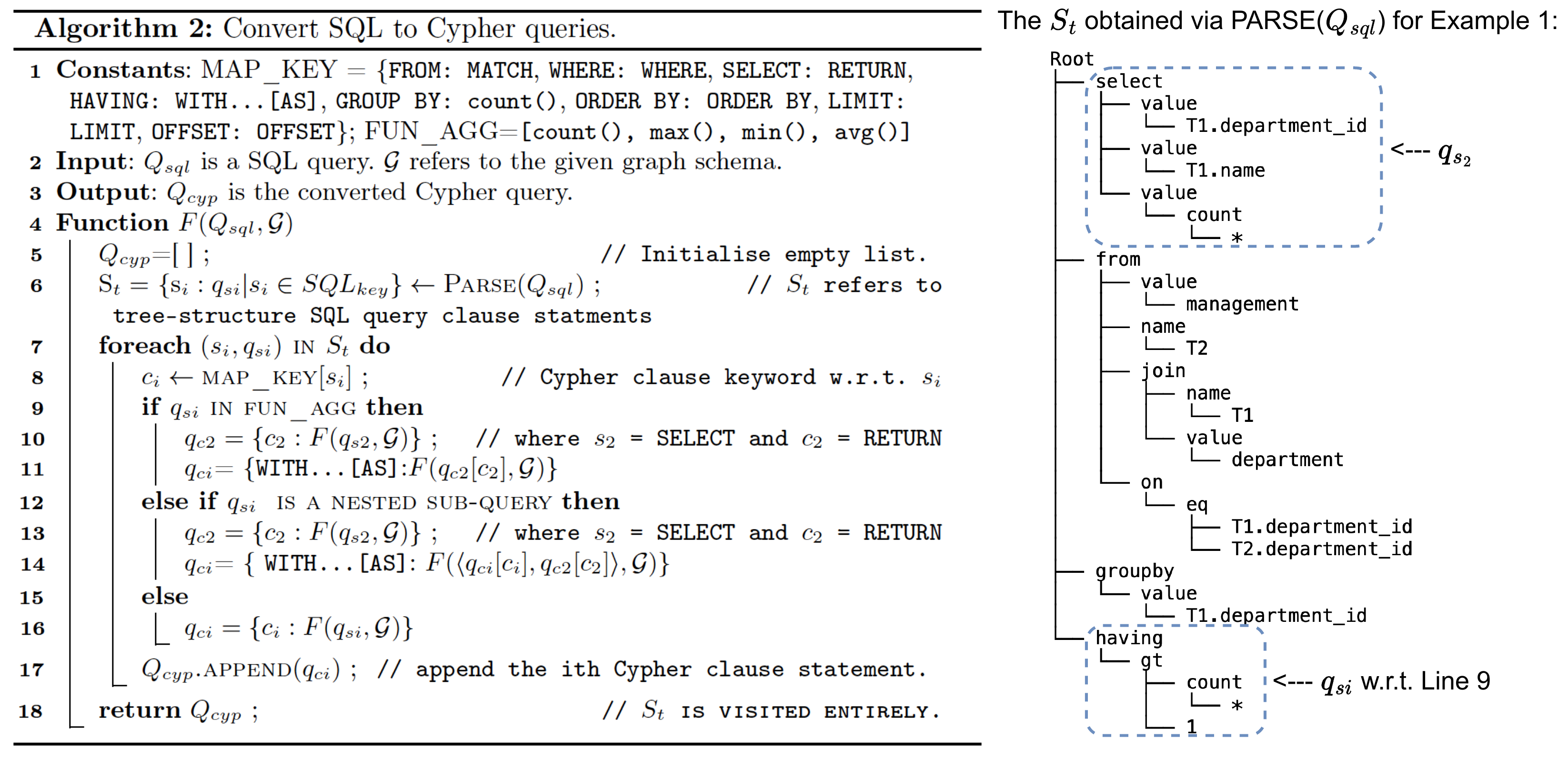}}
    \label{fig:algorithm}
\end{figure*}

Therefore, the essential translation tasks for SQL2Cypher are to sort out the equivalent mapping between SQL and Cypher clauses and identify graph patterns. We denote a list of SQL keywords as {\tt $SQL_{key}$}, i.e., {\tt [FROM, WHERE, SELECT, HAVING, GROUP BY, ORDER BY, LIMIT, OFFSET, UNION]}, and a list of Cypher keywords as {\tt $Cypher_{key}$}, i.e., {\tt [MATCH, WHERE, RETURN, WITH...[AS], ORDER BY, LIMIT, OFFSET]}. We allign {\tt FROM} with {\tt MATCH}, {\tt SELECT} with {\tt RETURN}, {\tt HAVING} with {\tt WITH...[AS]}, {\tt GROUP BY} with {\tt count()}, and the rest keywords stay the same. We map {\tt JOIN} statements to graph patterns. We chain schema items occurring in nested sub-queries or group queries together by {\tt WITH} clause, piping the results from the nested query as the starting points of the parent query. The mapping process is repeated until all SQL clauses are visited. The pseudocode of SQL-to-Cypher process is described in Algorithm 1. We use $\textsc{parse}(Q_{sql})$\footnote{https://github.com/mozilla/moz-sql-parser} to parse a SQL query into a $\textsc{JSON-izable}$ parse tree (Line 6), where $s_i$ refers to a SQL keyword appearing in $SQL_{key}$ list and each $s_i$ corresponds to its equivalent Cypher keyword $c_i$ in $Cypher_{key}$ list. The mapping between $s_i$ and $c_i$ becomes a $\textsc{KEY-VALUE}$ pair of $\textsc{MAP\_KEY}$ (Line 1). $q_{si}$ is the $\textsc{JSON-izable}$ parse tree w.r.t. $s_i$. The parsed tree hierarchy is illustrated next to Algorithm 1. Given Example 1 in Figure~\ref{fig:examples}, for Line 9 in Algorithm 1, $q_{si}$ with respect to {\tt HAVING} is shown in the dotted box at the bottom right of the parsed tree.

The mapping algorithm presented in Algorithm 2 has been meticulously designed to address the unique challenges and advantages of translating SQL queries into Cypher queries for graph databases. Several key considerations drove the design choices:

1) {\bf Efficient Handling of Highly Connected Data}. Graph data models excel in handling highly connected data, eliminating the need for multiple costly {\tt JOIN ON} operations used in SQL queries upon relational databases. Therefore, the algorithm is designed to seamlessly translate SQL queries into Cypher patterns, taking advantage of the graph data model's inherent ability to represent complex relationships without the explicit use of foreign keys. As the relationships between nodes are explicitly represented through edges. Thus, the algorithm is designed to omit unnecessary references to foreign keys in the translated Cypher queries, resulting in more concise and efficient queries.

2) {\bf Case Sensitivity Compatibility} Cypher queries are case-sensitive, whereas SQL queries may not be. To ensure compatibility, the algorithm normalizes all schema items from the source SQL queries to adhere to Cypher's case sensitivity, providing consistent and accurate evaluation.

3) {\bf Structured Mapping Process}. The algorithm's structured mapping process systematically aligns SQL keywords with their corresponding Cypher counterparts, simplifying the translation and ensuring that all clauses are processed methodically.

4) {\bf Handling Nested Queries and Grouping}. To address the complexity of nested sub-queries or group queries, the algorithm employs the {\tt WITH} clause to chain schema items, ensuring that results from nested queries serve as starting points for parent queries. This approach simplifies the translation of complex SQL structures into Cypher patterns.

By designing the mapping algorithm in this manner, we capitalize on the strengths of graph data models, minimize redundancy, and provide a systematic and efficient translation process from SQL to Cypher. These design choices are essential to achieve accurate and performant SQL-to-Cypher translations, facilitating the integration of relational databases with graph databases.

\section{Mapping Evaluation}

\subsection{Datasets}
In order to make our proposed approach more general and promise to work out-of-the-box on other datasets and illustrate the effectiveness of the proposed method in real-world scenarios, we evaluated the method on the challenging Spider and the KaggelDBQA datasets. Table~\ref{tab:data-stat} shows the data statistics.

\begin{table*}[h!]
    \centering
    \scalebox{0.8}{
    \begin{tabular}{ccccc}
    \toprule
   {\bf Dataset}  & {\bf \#DB} & {\bf \#Table/DB} & {\bf \#Row/DB} & {\bf \#$Q_{sql}$} \\
   \midrule
{\bf SubSpider ($\mathcal{S}$)} &  155& 5.1 & 2K &5,918\\
  % {\bf sub-Spider ($\mathcal{S}$)} &  155& 5.1 & 2K &5,918  & 6,670 \\
  \midrule
 {\bf KaggleDBQA ($\mathcal{S}$)} &8 & 2.3 & 280K & 272   \\
  % {\bf KaggleDBQA ($\mathcal{S}$)} &8 & 2.3 & 280K & 272 &  52  \\
    \bottomrule
    \end{tabular}}
     \caption{The dataset statistics of SubSpider and KaggleDBQA. Note that if there is at least one table with a number of rows over 6,000, we filter out the whole relational database. This action resulted in a sub-Spider comprising 155 relational databases accompanied by 5,918 intricate SQL queries and simpler ones. } 
    \label{tab:data-stat}
\end{table*}

\paragraph{\bf Spider}

We investigate to map and evaluate a subset of Spider. Specifically, considering the computational capacity, we drop 11 DBs of which there is at least one table containing a threshold number of rows. The Spider comprising more than 10,000 data, is more appropriate for the development of models that prioritize SQL and database knowledge as opposed to the only memorization of domain knowledge or values. 

\paragraph{\bf KaggleDBQA} Despite the improvements on large-scale relational databases based benchmarks such as Spider, the majority of cross-domain benchmark datasets continue to focus on the database schema as opposed to the database values, thereby resulting in a large gap from the real-world scenarios. KaggleDBQA attempted to mitigate this issue by constructing 272 SQL queries from eight databases. The relational databases in KaggleDBQA are pulled from the open-source data analysis platform Kaggle and are not normalised. 

\subsection{Evaluations}

\begin{figure}[h]
  \centering
    \scalebox{1}{\includegraphics[width=\textwidth]{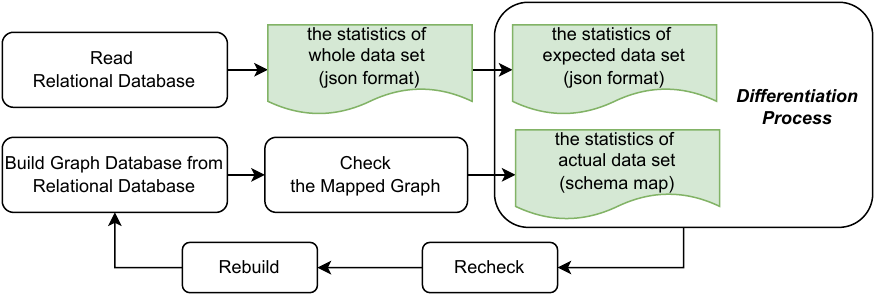}}
    \caption[]{Phases of the {\tt Check Mapping Consistency and Repair} Scenario. }
    \label{fig:check-repair-scenario}
\end{figure}

\paragraph{\bf Check Mapping Consistency and Repair}
We use the simple ``Repair'' scenario of the database mapping process (see Figure~\ref{fig:check-repair-scenario}) to evaluate mapping consistency in task 1. In this scenario, the expected data statistics are calculated. Secondly, the mapping result is loaded using {\tt APOC Procedures}, such as {\tt Apoc.meta.graph()}. Then, the statistics of the obtained graph are computed in terms of node labels, edge types, and properties. Next, a difference between the two is identified and revalidated. The mapping process is rechecked and the graph is rebuilt until there is no difference between the expected and actual database statistics.

\paragraph{\bf Execution Accuracy (EA)}
Execution accuracy ({\bf EA})~\citep{LinRX2020:BRIDGE} is employed to assess data consistency. This metric quantifies the proportion of information derived from Cypher queries executed on a graph database that aligns with the outcomes of SQL queries performed on relational databases. In the context of this evaluation, we denote the result set as $R_{cyp}^n$, representing the output of the $n^{th}$ Cypher query $Q_{cyp}$, while the result set $R_{sql}^n$ corresponds to the outcomes obtained from the corresponding SQL query $Q_{sql}$. The computation of EA is formulated as follows:

\begin{equation}
\small
    EA = \frac{\sum_{n=1}^{N} \mathbb{1}(R_{sql}^n, R_{cyp}^n)}{N} , where, \mathbb{1}(R_{sql}^n, R_{cyp}^n)= 
    \begin{cases}
        0 & \text{if } R_{sql}^n \neq R_{cyp}^n \\
        1 & \text{if } R_{sql}^n =  R_{cyp}^n \\
    \end{cases}
\end{equation}

where $R_{sql}^n$ denotes the result set by executing the $Q_{sql}^n$ query, and $R_{cyp} ^n$ denotes the result set by executing the semantically equivalent Cypher query $Q_{cyp}^n$.

\paragraph{\bf Valid Score (VS)}
VS is designed to measure the percentage of the cases where SQLs are successfully parsed into JSON-serializable parse trees using an openly available tool\footnote{https://github.com/mozilla/moz-sql-parser}, and accurately translated into the corresponding Cypher query via our proposed Algorithm 2. Any SQL queries that fail to be parsed into \textsc{JSON} documents will be declared invalid in our context. Compared with the EA metric, the VS could be considered a strict evaluation metric since it also takes into account the reliability of a SQL parser. Formally, the VS can be expressed as:

\begin{equation}
\small
    EA = \frac{\sum_{n=1}^{N} \mathbb{1}(R_{sql}^n, R_{cyp}^n)}{N+N(F_{sql})} , where, \mathbb{1}(R_{sql}^n, R_{cyp}^n)= 
    \begin{cases}
        0 & \text{if } R_{sql}^n \neq R_{cyp}^n \\
        1 & \text{if } R_{sql}^n =  R_{cyp}^n \\
    \end{cases}
\end{equation}
where $R_{sql}^n$ denotes the result set by executing the $Q_{sql}^n$ query, and $R_{cyp} ^n$ denotes the result set by executing the semantically equivalent Cypher query $Q_{cyp}^n$. $N(F_{sql})$ represents the count of cases in which an SQL parser failed to generate \textsc{JSON} documents.

\subsection{Results and Analysis}
In Table~\ref{tab:graph}, we present the overall statistics of the mapped datasets, including the statistics of the obtained graph databases and the corresponding Cypher queries. The table provides an overview of the graph database generated from the relational databases of SubSpider and KaggleDBQA, including the number of domains (\#DB), the count of graph nodes (\#Nodes), the total number of edges (\#Edges), and the number of Cypher queries that produced accurate results. 

\begin{table*}[h!]
    \centering
    \scalebox{0.8}{
    \begin{tabular}{lccccc}
    \toprule
    {\bf Property KG}&  {\bf \#Nodes}  & {\bf \#Edges}  & {\bf \#$Q_{correct\_cyp}$}  & {\bf \#$Q_{total\_cyp}$} & {\bf EA (\%)} \\
   \midrule
  {\bf CySpider} &  10,729  & 11,277  & 4,929 & 5574 & 88.43 \\
  \midrule
  {\bf CyKaggleDBQA} & 4,760,601 & 2,560,961 & 173 & 272 & 63.60  \\
    \bottomrule
    \end{tabular}}
     \caption{The statistics of the property knowledge graphs CySpider mapped from SubSpider and CyKaggleDBQA mapped from KaggleDBQA. EA shows the data consistency accuracy of the dataset mapping process. } 
    \label{tab:graph}
% \vspace{-2em}
\end{table*}

Following the definitions and categorizations proposed by J Marton et al.~\citep{marton2017formalising}, we focus on the statistics of Cypher queries as follows.

\begin{table}[h]
	\centering
	\resizebox{\textwidth}{!}
	{\begin{tabular}{r|r|r|r|r|r}
     \toprule
     \multicolumn{6}{c}{\bf CySpider Part 1: Graph database Statistics} \\
    \midrule
    \multicolumn{3}{c|}{\bf Graph Node} & \multicolumn{3}{c}{\bf Graph Edges} \\
    \cmidrule(lr){1-3} \cmidrule(lr){4-6} 
     {\bf \# DB } & {\bf \# Labels }& {\bf Nodes Count } 	 & {\bf \# DB } & {\bf \#  Types } & {\bf  Edge Count }\\
    \midrule
    155 & 641 & 10,729   & 141 & 461 & 11,277\\
    \midrule
    % \midrule
\multirow{2}{*}{\bf Split}  & \multicolumn{5}{c}{\bf CySpider Part 2: Cypher Queries Statistics}   \\ \cmidrule(lr){2-6}
     & \bf  \# Node Patterns & \bf \# Edge Patterns  & \bf \# Filtering Patterns & \bf \# Negation patterns  & \bf \# Aggregation\\
     \midrule
    Train & 3,046 & 1,684  & 2,401 & 113 & 1,739 \\
    Development & 407 & 298 & 342 & 20 &  321\\
\midrule
\midrule
     \multicolumn{6}{c}{\bf CyKaggleDBQA Part 1: Graph database Statistics} \\
    \midrule
    \multicolumn{3}{c|}{\bf Graph Node} & \multicolumn{3}{c}{\bf Graph Edges} \\
    \cmidrule(lr){1-3} \cmidrule(lr){4-6} 
     {\bf \# DB } & {\bf \# Labels }& {\bf Nodes Count } 	 & {\bf \# DB } & {\bf \#  Types } & {\bf  Edge Count }\\
    \midrule
    8 & 17  & 4,760,601 & 4 & 8 & 2,560,961 \\
    \midrule
    % \midrule
\multirow{2}{*}{\bf Split}  & \multicolumn{5}{c}{\bf CyKaggleDBQA Part 2: Cypher Queries Statistics}   \\ \cmidrule(lr){2-6}
     & \bf  \# Node Patterns & \bf \# Edge Patterns  & \bf \# Filtering Patterns & \bf \# Negation patterns  & \bf \# Aggregation\\
     \midrule
    GeoNuclearData & 22 &  0 & 18 & 0 & 13\\
    GreaterManchesterCrime & 24 & 0 &  18& 0 & 21 \\
    Pesticide &30 & 1 & 14 &0& 18 \\
    StudentMathScore & 14 & 1 & 9 & 0 & 2 \\
    TheHistoryofBaseball & 19 & 5 & 16 &0 &8\\
    USWildFires & 26& 0 & 10 & 0& 22 \\
    WhatCDHipHop &22 & 0 &9&0 & 3 \\
    WorldSoccerDataBase & 13& 0 & 9 & 0 & 7 \\
\bottomrule
    \end{tabular}}
    \caption[]{Statistics of CySpider and CyKaggleDBQA, including the graph database statistics and the corresponding Cypher queries, constructed from a subSpider and KaggleDBQA respectively.}
    \label{tab:rel2graph_stat}
\end{table}

\begin{itemize}
    \item \textit{Node and edge patterns} are the basic parts which are written in {\tt MATCH} clause. In this paper, we filter node and edge patterns in {\tt WHERE} clause rather than providing node label/edge type constraints in {\tt MATCH} clause. 
    \item \textit{Filtering patterns} express the uniqueness criterion for nodes and edges in a compact way, such as {\tt WHERE} clause.
    \item An example of \textit{negation patterns} is demonstrated by example 2 in Figure~\ref{fig:examples}. 
    \item We handle a subset of Cypher \textit{Aggregation} operators, i.e. {\tt count}, {\tt avg}, {\tt max}, {\tt min}, {\tt sum}. Aggregation operators often go together with {\tt WITH} clauses to realize grouping (see example 1 in Figure~\ref{fig:examples}). 
    \item There are 71 examples containing {\tt UNION} clause. Using just {\tt UNION} will combine and remove duplicates from the result set.
\end{itemize}

Table~\ref{tab:rel2graph_stat} showcases the statistics of the obtained counterpart dataset in detail. The table is divided into two sections, each providing the key elements of the corresponding graph database and the correctly executed Cypher queries mapped from the SQL queries. The part one sections offer the overviews of the CySpider and CyKaggleDBQA graph databases' statistical data, encompassing details on the number of databases, as well as the labels and types attributed to nodes and edges. Part two sections delve into the statistics related to Cypher queries. This part includes data on node patterns and edge patterns, filtering patterns, negation patterns, and statements that incorporate aggregation operators.

\paragraph{\bf Mapping Analysis}

Table~\ref{tab:graph} shows that the EA score of CyKaggleDBQA is lower than that of CySpider, despite CySpider's graph being significantly smaller than CyKaggleDBQA's graph. Additionally, the number of SQL query variants in KaggleDBQA is less than that in CySpider. Hence, the EA score demonstrates sensitivity to both the size of the constructed graph and the diversity of SQL query variants. We may consider exploring alternative evaluation metrics that can provide a more comprehensive assessment of our mapping results. 

\begin{figure*}[h]
  \centering
    \scalebox{1}{\includegraphics[width=\textwidth]{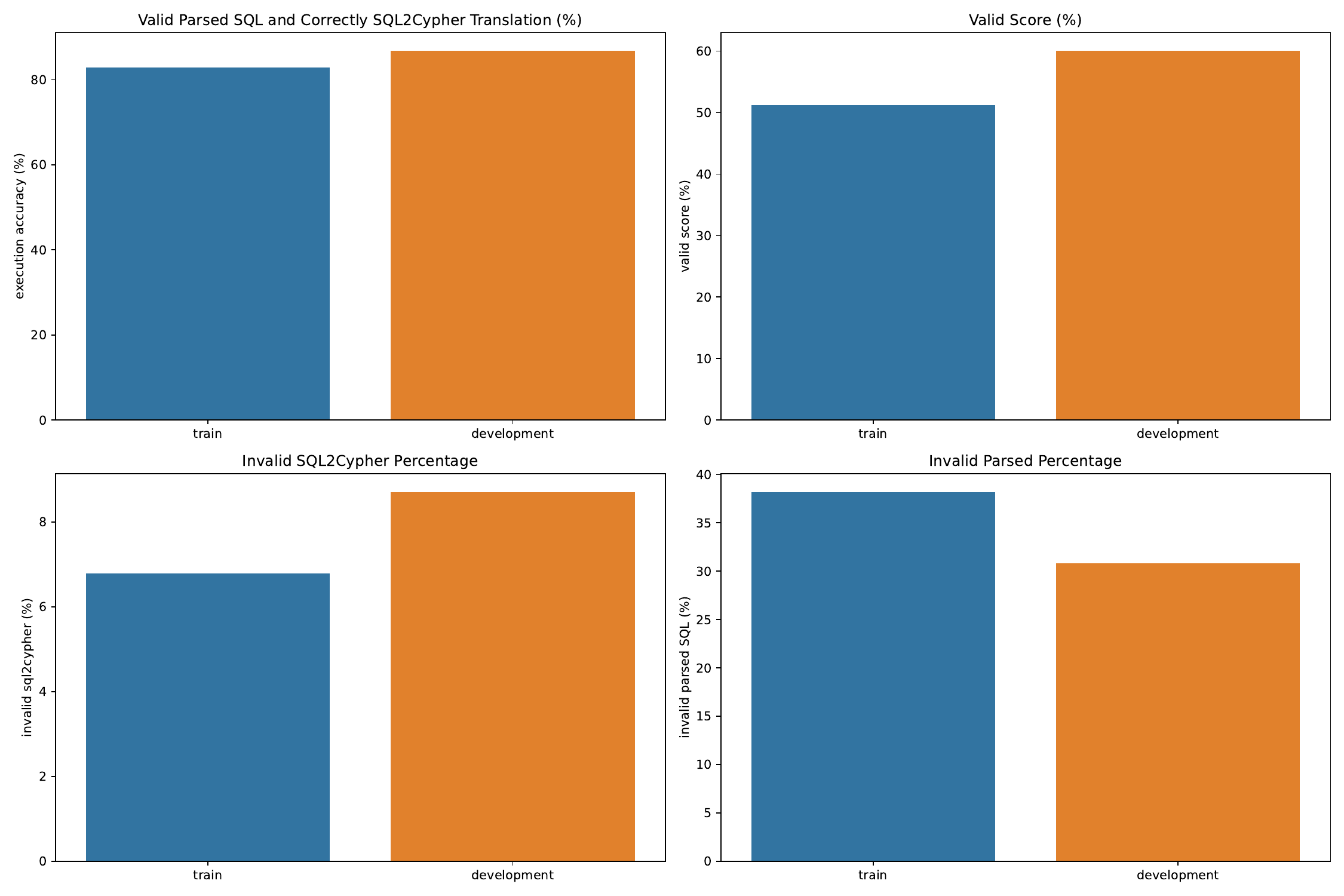}}
    \caption{The query mapping quality distribution from the four aspects on SubSpider. }
    \label{fig:subspider_query_mapping}
\end{figure*}

\begin{figure*}[h]
  \centering
    \scalebox{1}{\includegraphics[width=\textwidth]{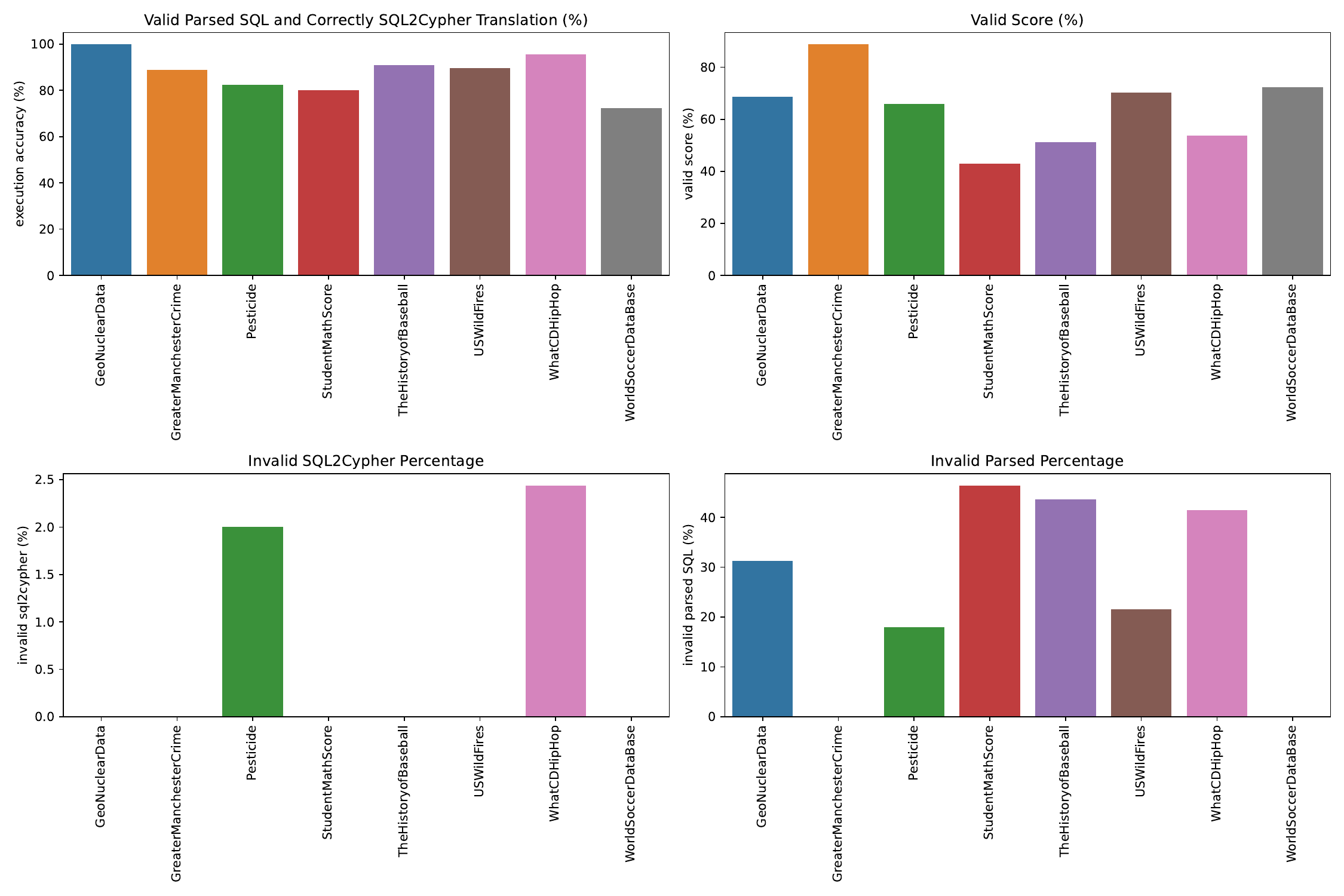}}
    \caption{The query mapping quality distribution from the four aspects on KaggleDBQA. }
    \label{fig:kaggle_query_mapping}
\end{figure*}

Our query mapping module initiates by parsing a SQL query into a JSON-serializable parse tree using an openly available tool\footnote{https://github.com/mozilla/moz-sql-parser}. Figures~\ref{fig:subspider_query_mapping} and~\ref{fig:kaggle_query_mapping} illustrate the EA, VS, invalid SQL2Cypher percentage and invalid parsed percentage from top to bottom and left to right. The VS score and invalid parsed percentage show inverse proportionality characteristics. The lower proportion of invalid parsed SQL contributed to the higher overall valid score, which suggests that a better SQL parser would enhance query mapping accuracy to Cypher queries. The development of an improved SQL parser would be one of the future works. A SQL parser with higher performance would facilitate the better translation of SQL to other database query languages such as SPARQL. 

Certain false negative cases arose due to limitations inherent in the Cypher query language. For instance, in the {\tt GreaterManchesterCrime} domain, all the incorrect execution results are owing to the instances of false negatives. These occur when the {\tt GROUP BY Location} statement in the source SQL queries implicitly constrains the relational database engine to sort the field in alphabetical order if it is of text type. However, this cannot be accurately translated into a Cypher clause.

\section{Conclusion}

This paper proposes an automated approach to migrate data and queries from relational databases to a property graph database. The translation leverages the integrity constraints defined over the source to systematically build a corresponding property knowledge graph. The primary aim of this approach was to enhance the inherent structure of relational data through graph representation, resulting in optimized query operations when addressing semantic-level tasks, e.g. semantic parsing. We use the execution accuracy metric to evaluate the mapping process at the formal semantic query language level including complex queries. We suggested evaluating our approach on another large-scale dataset such as BIRD~\citep{li2023can} to further enhance the mapping process’s accuracy, efficiency, and coverage for comparable experimental outcomes.

\section*{Acknowledgment}
This research is supported by the Australian Research Council through the Centre for Transforming Maintenance through Data Science (grant number IC180100030), funded by the Australian Government.

\bibliographystyle{unsrtnat}
\bibliography{arxiv}

\end{document}